%% file: LaponiteConvection.tex
\documentclass[epj]{svjour} 

\usepackage{amsmath}
\usepackage{amssymb}
\usepackage{psfrag}
\usepackage{graphicx}
\usepackage{color}
\usepackage{sistyle} 

\definecolor{orange}{rgb}{1.0,0.7,0.0}

\input{matlabfig.tex}

\begin{document}

\title{\bf Coupling between aging and convective motion in a colloidal glass of Laponite}

\author{L. Bellon\thanks{\emph{Present address:} \'Ecole Normale Sup\'erieure de Lyon, Laboratoire de Physique, 46, All\'ee d'Italie, 69364 Lyon Cedex 07, France} \and M. Gibert\footnotemark[\value{footnote}] \and R. Hern\'andez}

\institute{LEAF-NL, Departamento de Ingeniera Mecanica, Universidad de Chile, Beaucheff 850, Santiago, Chile}

\mail{\tt Ludovic.Bellon@ens-lyon.fr}

\date{Final draft of Eur. Phys. J. B {\bf 55}, 101--107 (2007). \\
Published online 15 February 2007 -- \copyright \ EDP Sciences, Societ\`a Italiana di Fisica, Springer-Verlag 2007.\\
The original publication is available at www.springerlink.com}

\abstract{We study thermal convection in a colloidal glass of Laponite in formation. Low concentration preparation are submitted to destabilizing vertical temperature gradient, and present a gradual transition from a turbulent convective state to a steady conductive state as their viscosity increases. The time spent under convection is found to depend strongly on sample concentration, decreasing exponentially with mass fraction of colloidal particles. Moreover, at fixed concentration, it also depends slightly on the pattern selected by the Rayleigh B\'enard instability: more rolls maintain the convection state longer. This behavior can be interpreted with recent theoretical approaches of soft glassy material rheology.
} 

\maketitle

\section{Introduction}\label{Part-intro}

Many systems in nature, such as glasses, spin-glasses, colloids and granular materials, present a very slow relaxation towards equilibrium and any physical property of these materials evolves as a function of time: they are aging. For example, when a glassy system is quenched from above its glass transition temperature $T_g$ to a temperature lower than $T_g$, any response function of the material depends on the time $t_w$ elapsed from the quench \cite{Struick}. Another example of aging is given by colloidal glasses, whose properties evolve during the sol-gel transition which may last several months \cite{Kroon}.

Although great experimental and theoretical efforts have been paid for years to understand the physics of glassy systems, a global coherent picture is still laking. Some recent theories \cite{Kurchan} based on the description of spin glasses by a mean field approach proposed to extend the concept of thermodynamic temperature using a Fluctuation Dissipation Relation (FDR) which generalizes the Fluctuation Dissipation Theorem (FDT) for a weakly out of equilibrium system (for a review see ref. \cite{Mezard,Cugliandolo,Peliti}). This novel approach has been quite successful, and is already illustrated by a few experiments in real materials \cite{Grigera,Bellon,BellonD,Herisson,Buisson}, where FDR could be measured and compared to theoretical predictions. The main difficulty of these experiments is the study of intrinsic fluctuations, since no long-time averaging can be done in an aging system.

It has been proposed that when a gentle forcing is applied to these systems, they could reach a stationary state where no aging can be observed, although they are still out of equilibrium \cite{Kurchan,Berthier}. In this state, the fluctuation-dissipation ratio would be similar to that of the glassy phase, with a two temperatures scenario in the case of p-spins glasses for example\cite{Berthier}. This theoretical idea is very appealing to the experimentalist as it would ease the fluctuation measurements, allowing long time averaging in a stationary system. The drawback of typical methods is that forcing the system, by adding a shear force for example, induces great perturbations on the fluctuation measurements due to unwanted vibrations, electrical disturbances, etc. inevitably produced by the forcing method.

Thermally driven flows may present a bypass for these problems: the classic Rayleigh B\'enard instability for example, can be used as the forcing method. Heating the sample from below, one triggers convective motion, and therefore a small shear associated to the flow \cite{Koschmieder}. The advantage of this method is that it does not imply any moving parts or motors, which would inevitably produce additional fluctuations.

In the study of soft glassy material, Laponite \cite{Laponite} has often been use as a standard example of colloidal glass, presenting many similarities with structural glass aging \cite{Kroon}. Rheology \cite{BellonD,Mourchid,Bonn,Bonn2002}, light and X-Ray diffusion \cite{Kroon,Mourchid,Bonn,Knaebel}, electrical measurements \cite{Bellon,BellonD} have been used to characterize aging of this material, and evidenced many interesting aspect of its glassy behavior. Nevertheless, no attention has been paid up to now to its thermal properties or to its behavior in a convective flow. We propose here a investigation in this direction, with the following questions in mind: are thermal properties of Laponite aging too ? Can convective flow modify (delay, accelerate, stop) aging behavior ? Would it make a suitable driving method to study the fluctuation-dissipation ratio of a forced glassy system ?

This article is organized as follow: we first describe the experimental set-up (Sect. \ref{Part-setup}), then study the basic properties relevant for convection in Laponite (Sect. \ref{Part-properties}). Sect. \ref{Part-convection} details the experimental results of convection in Laponite, before the concluding section (Sect. \ref{Part-conclusion}).

\section{Experimental setup} \label{Part-setup}

\begin{figure}[!ht] 
	\begin{center} 
		\psfrag{T}[Bl][Bl]{\scriptsize Temperature sensors (Pt100)}  
		\psfrag{H}[Bc][Bc]{\scriptsize Hot plate (heater foil)}  
		\psfrag{C}[Bl][Bl]{\scriptsize Cold plate (water circulation)}  
		\psfrag{L}[Bc][Bc]{\tiny $L$}  
		\psfrag{p}[Bc][Bc]{\tiny $w$}  
		\psfrag{h}[Bc][Bc]{\tiny $h$}  
		\psfrag{x}[Bc][Bc]{\tiny $x$}  
		\psfrag{y}[Bc][Bc]{\tiny $y$}  
		\psfrag{z}[Bc][Bc]{\tiny $z$}  
		\psfrag{g}[Bc][Bc]{\tiny $\mathbf{g}$}
	
		\includegraphics[width=8cm]{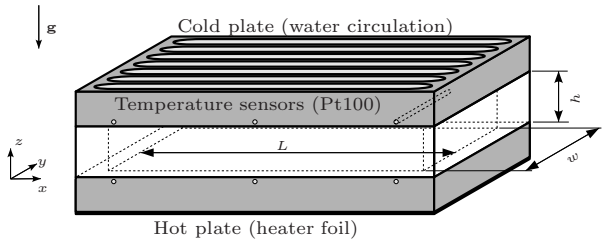}
	\end{center}
\caption{Experimental setup: the convection cell is rectangular with aspect ratios close to $6:1$ and $3:1$ ($L=\SI{145}{mm}$, $w=\SI{68}{mm}$, $h=\SI{23,5}{mm}$). The cold plate temperature is regulated with a water circulation, while the other plate is heated with a flat sheet resistance. Several Pt100 sensors are used to measure temperature stability and uniformity.}  \label{Fig-setup}
\end{figure}

\subsection{Convection cell}

The experimental setup is illustrated in Fig.\ref{Fig-setup}. The convection cell is rectangular: ($L \times w \times h$) $\SI{145}{mm}$ long, $\SI{68}{mm}$ deep and $\SI{23.5}{mm}$ high. The lateral walls are made of $\SI{10}{mm}$ thick plexiglas, and top/bottom plates are made of $\SI{16}{mm}$ thick stainless steel\footnote{Although stainless steel has a poor thermal conductivity with respect to the other metals usually used in a convection cell, it is still more than one order of magnitude more conductive than water and ensures chemical neutrality in Laponite preparation aging.}. Several platinum sensors (RTD-100) are placed in the two metal plate bodies, close to the cell surface, to measure temperature gradient and uniformity. All measurements are made with a Keithley 2000 multimeter and scan card in 4 wires configuration. The cold plate temperature is controlled with a circulation from a water bath (Nestlab RTE-210), allowing better than $\SI{0.1}{K}$ temperature uniformity and stability. The other plate is heated with a $\SI{18}{W}$ flat sheet resistance of matched size. This resistance is connected to a DC power supply (HP E3632A) in a 4 wires configuration, which allows precise heating power setting and monitoring. All instruments are remote controlled with a computer via GPIB interface, which allows sampling temperature of the various sensors at $\SI{0.1}{Hz}$ and controlling hot temperature with a PID algorithm: we achieve better than $\SI{0.5}{K}$ temperature uniformity and $\SI{0.03}{K}$ stability. The temperature of the external environment is stabilized to better than $\SI{0.5}{K}$ over $\SI{24}{h}$. To minimize heat exchange of the cell with the environment, the hot and cold plate temperature are set symmetrically to the room one, and both plates are imbedded in a tick polystyrene insulation. In the worst case (conductive configuration), heat losses represent $\SI{30}{\%}$ of total heating power.

\begin{figure*}[!ht]\unitlength=1mm
	\begin{center}
	\begin{picture}(0,50)	\put(-75,0){	  
		\psfrag{Screen}[Br][Br]{\small Slanted screen}	  
		\psfrag{Camera}[Br][Br]{\small Digital camera}	  
		\psfrag{Light}[Bl][Bl]{\small Laser light sheet}	  
		\psfrag{Cell}[Bl][Bl]{\small Convection cell}
		\includegraphics[width=8cm]{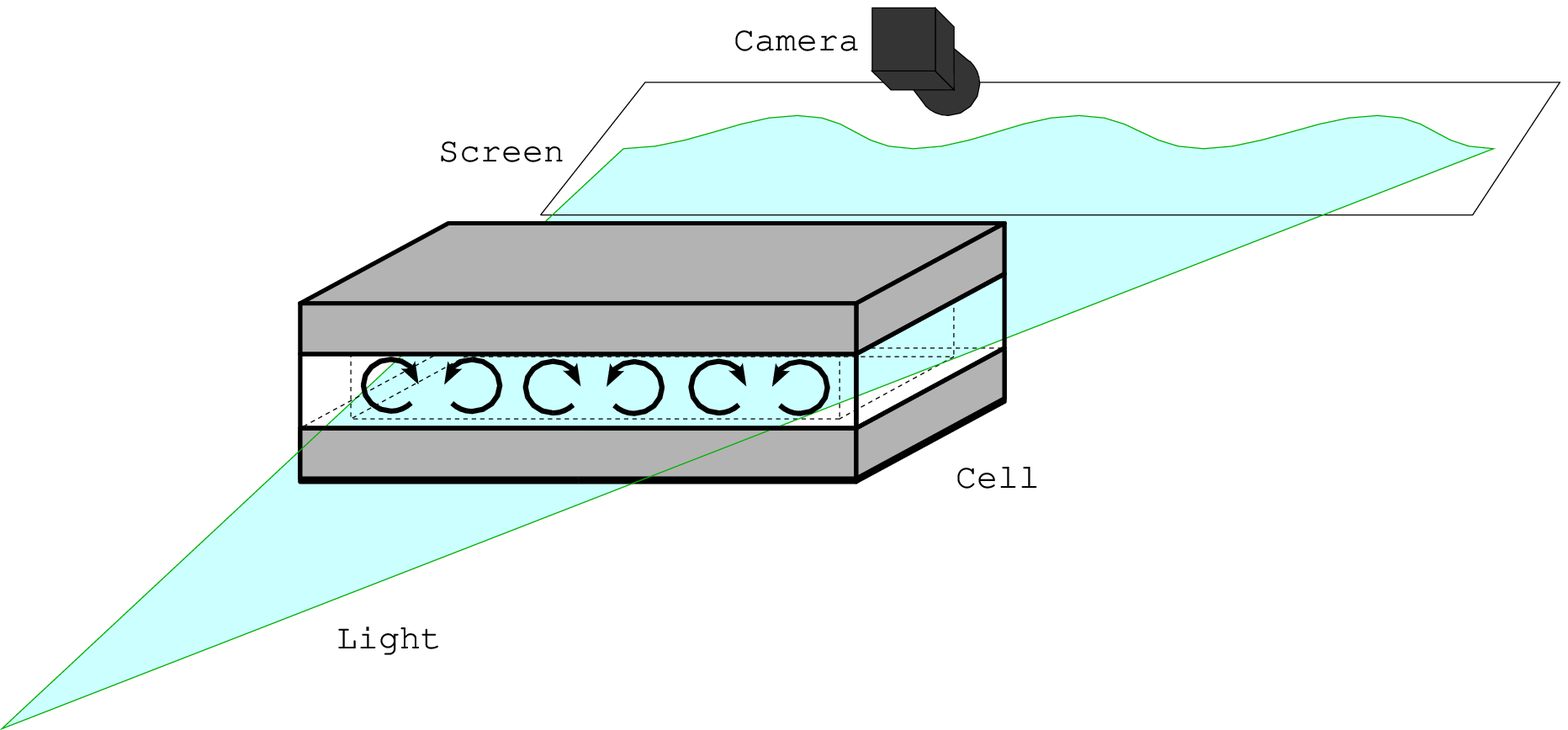}}
		\put(-50,40){\includegraphics[width=6cm]{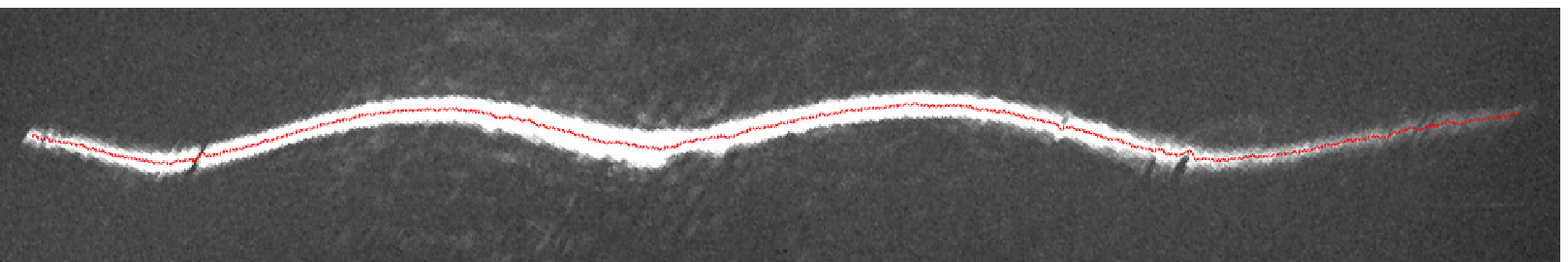}}	
		\put(10,25){
			\begin{minipage}{7cm}
				\small		
				\xbottomlabel[3mm]{Horizontal coordinate, $x/\SI{}{cm}$}		
				\yleftlabel[4mm]{$\partial_{\! z} \, T / \partial_{\! z} \,T_0$} 		
				\includematlabfig{5cm}{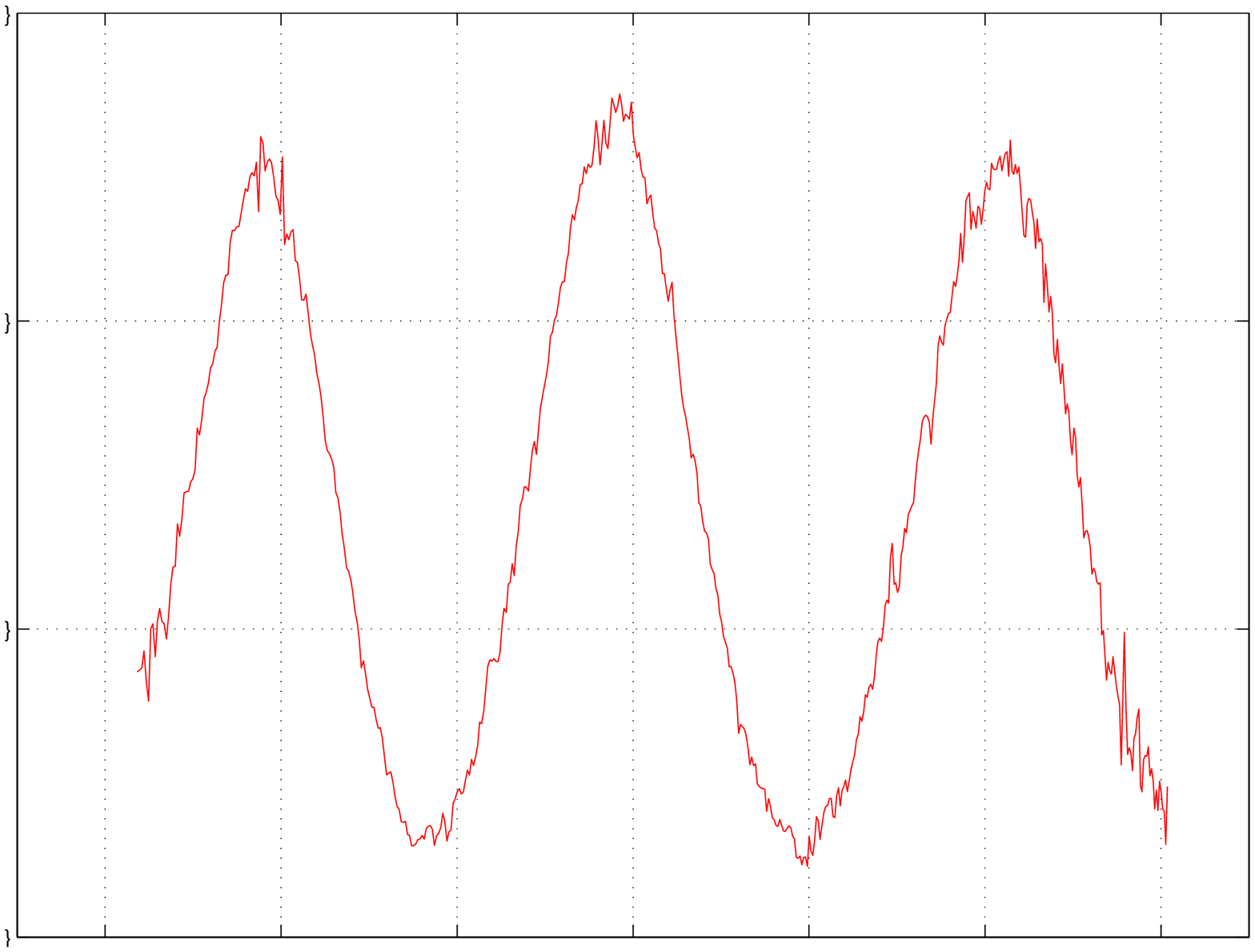}		
			\end{minipage}	}
	\end{picture}
	\end{center}
\caption{Pattern visualization: A horizontal laser light sheet crosses the cell along its shorter horizontal length, $\SI{2}{mm}$ above the bottom plate. It is deviated by the temperature gradient integrated along its path through the cell. Its projection on a slanted screen (to magnify deviation) follow the temperature pattern, and can be recorded with a digital camera. We extract from the raw image the position of the maximum intensity line (thin red line in the center of the light sheet image). After appropriate rescaling, we can plot on the right figure the vertical temperature gradient (normalized by the measurement corresponding to the conductive case) along the cell. We present here the example of 6 regular convection rolls.}  \label{Fig-visu}
\end{figure*}

\subsection{Pattern visualization}

To perform a visualization of the convective flow pattern, we propose here a very simple technique. We produce a horizontal laser light sheet using a cylindrical lens and a diode pumped solid state laser (Melles Griot 58GCS405, $\SI{5}{mW}$ output at $\SI{532}{nm}$). This light sheet crosses the cell along its shorter horizontal length, close to the bottom plate (at $z=\SI{2}{mm}$). The temperature gradient inside the cell induces an optical index gradient, mainly in the ($x$,$z$) plane, deviating the light from its straight path. The light sheet is thus modulated by the temperature pattern, and its projection on a screen can be monitored with a camera: the line obtained represent the average of the temperature gradient in the $y$ direction (shorter horizontal length of the cell). The slight horizontal deviation cannot be measured since it takes place inside the light plane, but the vertical deviation is easily recorded: for a typical temperature difference $\Delta T = \SI{8}{K}$ in the cell filled with water, computed and observed deviation is $\SI{2e-3}{rad}$, when beam pointing stability is better than $\SI{2e-5}{rad}$. Thus relative variation up to $\SI{1}{\%}$ can be detected on the vertical temperature gradient $\partial_z T$. 

This simple technique is sufficient to describe the flow pattern main characteristics such as size and periodicity of spatial structures, time evolution, etc. Quantitative measurement of the vertical temperature gradient can even be made when the deviation is compared to reference situations (zero and uniform temperature gradient). Eventually, performing a systematic vertical scanning would allow the complete determination of temperature field, but this technique is beyond the scope of the present work. We present in figure \ref{Fig-visu} an example of this visualization technique in the case of 6 steady convection rolls.

\subsection{Laponite preparation}

A great attention is paid to prepare the Laponite samples always in the same way, minimizing the source of non reproducibility. Preparation are done in a clean $\mathrm{N}_2$ atmosphere, to avoid $\mathrm{CO}_2$ contamination. Laponite RD powder is mixed in a given mass fraction with distillate water for $\SI{1}{h}$ under strong magnetic stirring. Concentration $C$ ranges from $\SI{(1.8}{}$ to $\SI{2.7)}{wt\%}$ in these experiments, leading to the slow formation of a colloidal glass in times ranging from months to hours. The solution, initially fairly liquid, is injected in the cell through a $\SI{1}{\micro m}$ pore filter (Whatman Puradisc 25 GD, $\SI{1}{\micro m}$ GMF-150) under $\SI{2}{bar}$ pressure. This event sets the initial time of the experiment. Filling and setting up operations take about $\SI{5}{min}$, which is considered as the time accuracy of our measurement. To minimize thermalization time of the hot (cold) plate, it is preheated (precooled) to the desired temperature before filling. Thanks to the symmetry of temperature limits around the room one, the mean temperature of the solution does not change and its volume is constant.

\section{Physical properties of Laponite samples} \label{Part-properties}

In a first set of experiments, we studied the time evolution of some physical properties relevant in Rayleigh B\'enard convection of Laponite preparations, in order to see if any aging behavior can be evidenced in these observables of the system, as for its viscous or electrical properties. First of all, we measured the thermal conductivity $\lambda$ and the thermal diffusivity $\kappa$ using our experimental setup in a conduction configuration: the cell is flipped upside down so that heating takes place from above, resulting in a vertical stabilizing temperature gradient, thus a conductive regime. The heat flux is imposed in a series of steps and the temperature difference across the cell is recorded for a $\SI{2.2}{wt\%}$ preparation of Laponite and for pure water as a calibration measurement. The transients in $\Delta T$ let us estimate the thermal diffusivity, while the stationary states lead to the conductivity. 
But as can be expected from their composition, none of the samples can be distinguished from pure water: no aging is observed and we measure \mbox{$\lambda=\SI{(0.6}{}\pm\SI{0.02)}{W.m^{-1}.K^{-1}}$} and $\kappa = \SI{(1,41}{} \pm \SI{0,07)e-7}{m^2.s^{-1}}$ at $\SI{25}{\degC}$.

Similarly, the thermal expansion coefficient $\alpha$ and density $\rho$ of Laponite preparation don't age and are equal to that of water within a few percent: for a $\SI{2.2}{wt\%}$ preparation at $\SI{25}{\degC}$, we measured $\rho= \SI{(1010}{} \pm \SI{3)}{kg.m^{-3}}$ and $\alpha = \SI{(2.0}{} \pm \SI{0.1)e-4}{K^{-1}}$. 

For a ideal newtonian fluid of kinemanic viscosity $\nu$, the control parameter of the Rayleigh B\'enard instability is the Rayleigh number, defined as
\begin{equation} \label{Eq-Ra}
Ra = \frac{g \alpha \Delta T h^3 }{\nu \kappa}
\end{equation}
where $g$ is the gravity constant and $h$ the height of the convection cell \cite{Koschmieder}. For our samples, the rheological behavior would thus be the only parameter presenting substantial aging and be responsible for the convective properties of Laponite preparations: convection should be present as long as the Rayleigh number is greater than the critical threshold $Ra_c$, that is as long as the viscosity is lower than the corresponding critical viscosity $\nu_c$. Using Laponite preparation properties, $\Delta T = \SI{8}{K}$ and assuming $Ra_c \approx \SI{2000}{}$ \cite{Koschmieder}, we compute $\nu_c \approx \SI{7.5e-4}{m^2.s^{-1}}$.

\begin{figure*}[!ht]
\unitlength=1mm
\begin{center}
\begin{picture}(0,63)
	\put(-90,32){
	\begin{minipage}{\columnwidth}			
		\xbottomlabel{Time, $t/\SI{}{h}$}			
		\yleftlabel[6mm]{Complex viscosity, $\nu^*/\SI{}{m^2.s^{-1}}$} 		  
		\psfrag{f=}[Bl][Bl]{\scriptsize $f\!=$}		  
		\psfrag{0.1Hz}[Bl][Bl]{\scriptsize \textcolor{red}{$\SI{0.1}{Hz}$}}		  
		\psfrag{0.3Hz}[Bl][Bl]{\scriptsize \textcolor{orange}{$\SI{0.3}{Hz}$}}		  
		\psfrag{1Hz}[Bl][Bl]{\scriptsize \textcolor{green}{$\SI{1}{Hz}$}}		  
		\psfrag{3Hz}[Bl][Bl]{\scriptsize \textcolor{cyan}{$\SI{3}{Hz}$}}		  
		\psfrag{10Hz}[Bl][Bl]{\scriptsize \textcolor{blue}{$\SI{10}{Hz}$}}		  
		\psfrag{nuc}[cc][cc]{\colorbox{white}{$\nu_c$}}
		\includematlabfig{7cm}{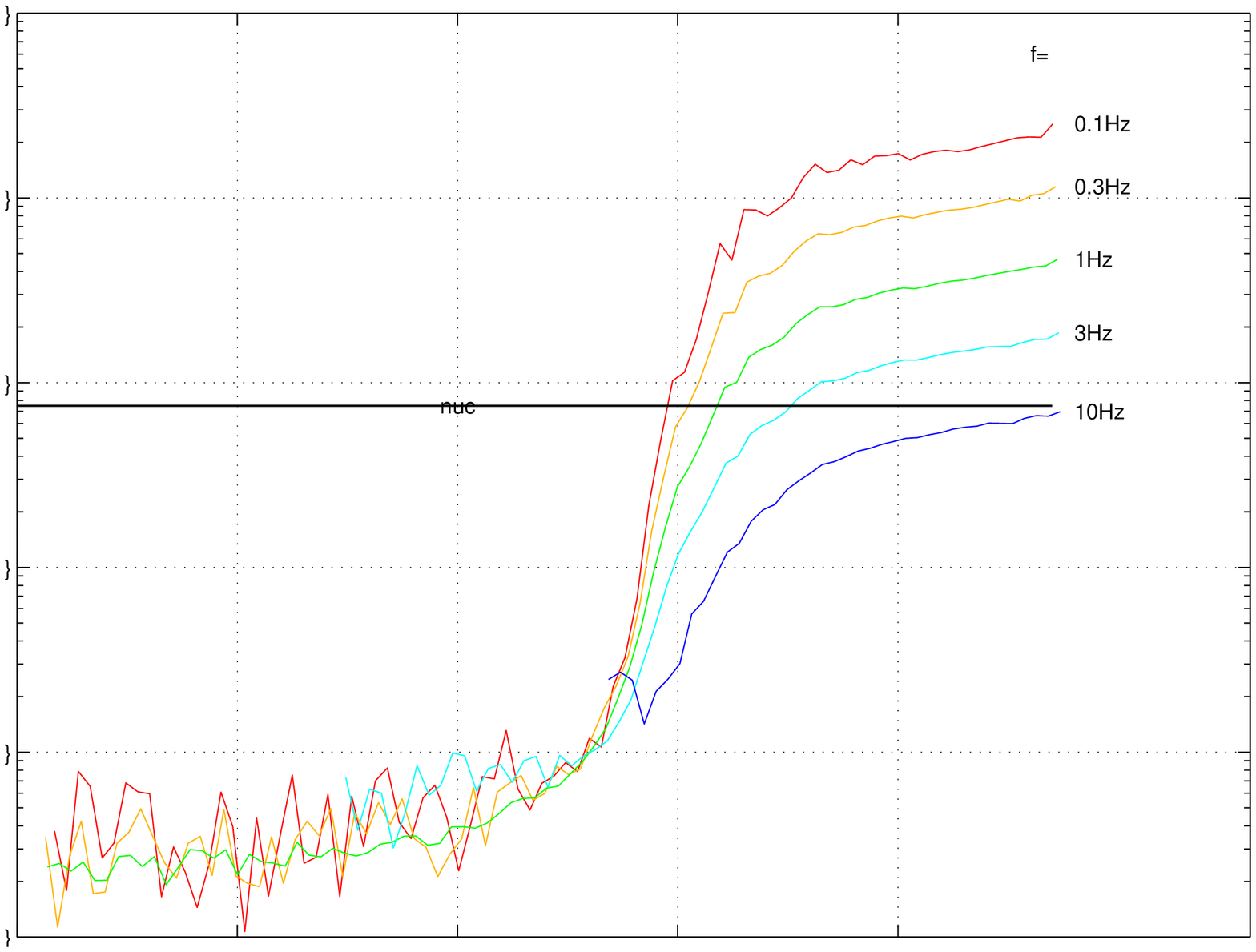}			
		\end{minipage}	}	
	\put(0,32){	
	\begin{minipage}{\columnwidth}			
		\xbottomlabel{Frequency, $f/\SI{}{Hz}$}
		\yleftlabel[6mm]{Complex viscosity, $\nu^*/\SI{}{m^2.s^{-1}}$} 		  
		\psfrag{t=}[Bl][Bl]{\scriptsize $t\!=$}		  
		\psfrag{f0.85}[Bl][Bl]{\colorbox{white}{\scriptsize $\propto f^{-0.85}$}}		  
		\psfrag{nuc}[cc][cc]{\colorbox{white}{$\nu_c$}}		  
		\psfrag{7h}[Br][Br]{\scriptsize \textcolor{red}{$\SI{7}{h}$}}		  
		\psfrag{12h}[Br][Br]{\scriptsize \textcolor{orange}{$\SI{12}{h}$}}		  
		\psfrag{14h}[Br][Br]{\scriptsize \textcolor{yellow}{$\SI{14}{h}$}}		  
		\psfrag{14.5h}[Br][Br]{\scriptsize \textcolor{green}{$\SI{14.5}{h}$}}		  
		\psfrag{15h}[Br][Br]{\scriptsize \textcolor{cyan}{$\SI{15}{h}$}}		  
		\psfrag{16h}[Br][Br]{\scriptsize \textcolor{blue}{$\SI{16}{h}$}}		  
		\psfrag{24h}[Br][Br]{\scriptsize \textcolor{magenta}{$\SI{24}{h}$}}
		\includematlabfig{7cm}{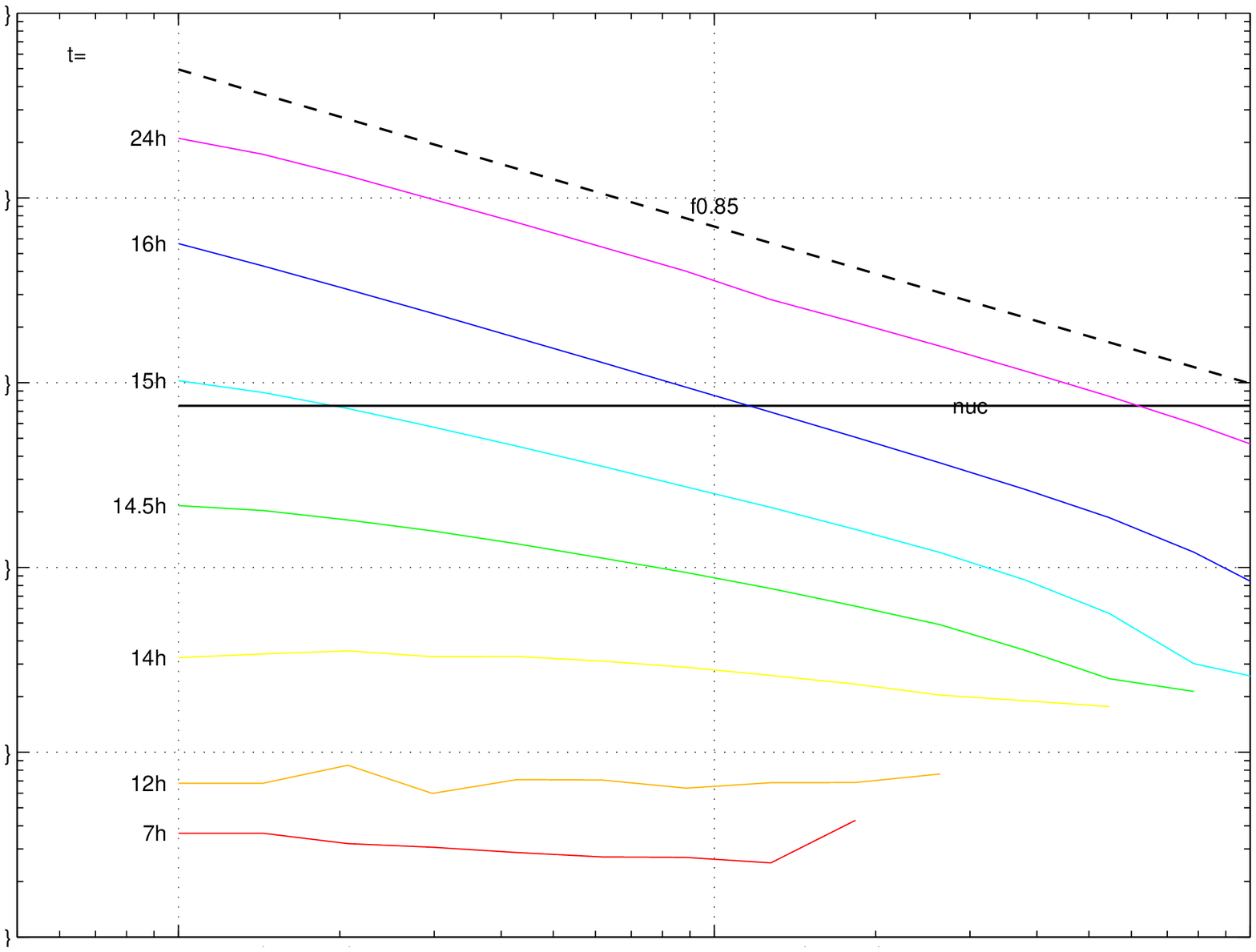}	
	\end{minipage}	}
\end{picture}
\end{center}
\caption{Complex viscosity of a $\SI{2.2}{wt\%}$ Laponite preparation. $\nu^*$ is plotted as a function of time for various frequencies (left), and as a function of frequency for different aging times (right). We observe a transition between an initial liquid state, where $\nu^*$ is roughly independent of $f$, and a solid like behavior for larger times, where $\nu^*$ scale as $f^{-0.85}$ (dashed line). Convection should occur as long as $\nu^*$ is smaller than $\nu_c$, thus about $\SI{15}{h}$ to $\SI{16}{h}$ at this specific concentration. Low time $/$ high frequency data is not available due to inertia limitations of the measurement. }  \label{Fig-nu}
\end{figure*}

We present in figure \ref{Fig-nu} a measurement of the complex viscosity $\nu^*$ of a $\SI{2.2}{wt\%}$ preparation at $\SI{25}{\degC}$. These results are obtained with a commercial Bohlin rheometer (C-VOR 150) in cone plane configuration ($\SI{60}{mm}$ diameter for $\ang{2}$ aperture), under a imposed strain of $\SI{5}{\%}$ and different frequencies $f$ ranging from $\SI{0.1}{Hz}$ to $\SI{10}{Hz}$. In this configuration, we test the linear viscoelastic response of the sample to harmonic shearing, and measure its elastic ($G^{\prime}$) and viscous ($G^{\prime\prime}$) moduli. The complex viscosity $\nu^*$ is then defined as
\begin{equation}
\nu^*=\frac{1}{\rho}\frac{\sqrt{G^{\prime2}+G^{\prime\prime2}}}{2 \pi f}
\end{equation}
For an ideal Newtonian liquid, this complex viscosity is independent of frequency and equal to the classic viscosity, whereas for a Hook solid, it scales as the inverse of frequency. The aging of Laponite rheological properties can be seen as a transition from a liquid like state (the viscosity being roughly independent of frequency and gradually increasing) to a solid like state, where $\nu^*$ decreases as $f^{-0.85}$ \cite{Bonn2002}. As seen on figure \ref{Fig-nu}, this transition occurs for aging times around $\SI{15}{h}$ to $\SI{16}{h}$ for a $\SI{2.2}{wt\%}$ preparation, when $\nu^*$ abruptly crosses the $\nu_c$ threshold at lower frequencies. As the apparition of the viscoelastic properties of Laponite preparations occurs lately, when the viscosity is already quite high, the convective behavior of the sample can be considered as that of a casual liquid whose viscosity increases with times. The elastic component might only play a role in the last part of the experiment, just before the onset of the conductive state. 

\section{Convection with Laponite} \label{Part-convection}

\newsavebox{\inset}\sbox{\inset}{
\begin{minipage}{5cm}
	\footnotesize
	\xbottomlabel[2mm]{$t/t_c$}
	\yleftlabel[2mm]{$Nu$} 
	\includematlabfig{2.5cm}{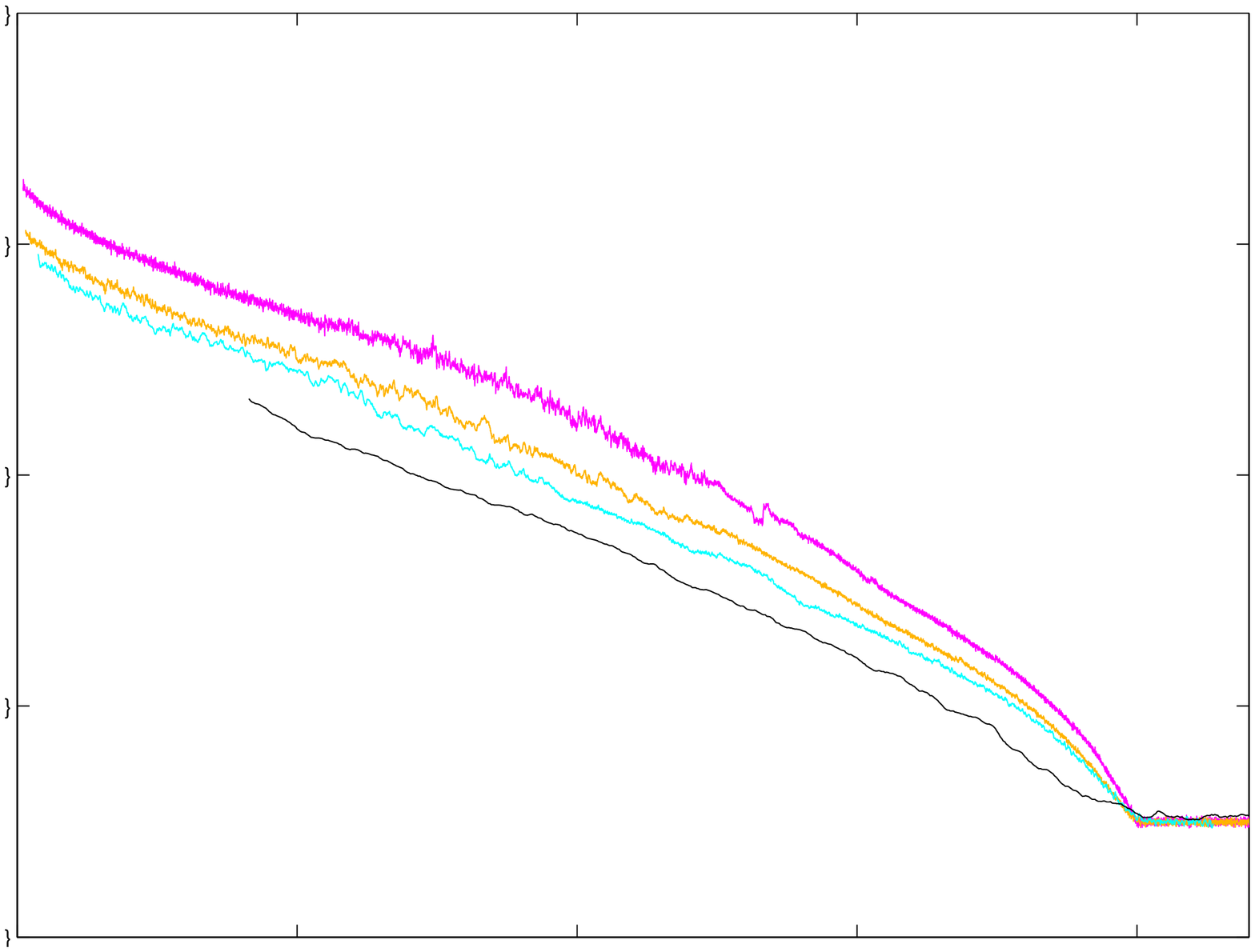}
\end{minipage}}

\begin{figure}[!ht]
\unitlength=1mm
\begin{center}
\begin{picture}(0,62)	
	\put(-45,32){
	\begin{minipage}{\columnwidth}
		\xbottomlabel{Time, $t/\SI{}{h}$}
		\yleftlabel[2mm]{Nusselt number, $Nu$} 
		\psfrag{C=}[Bl][Bl]{\small $C\!=$}		  
		\psfrag{2.0}[Bl][Bl]{\small \textcolor{red}{$2.0$}}		  
		\psfrag{2.05}[Bl][Bl]{\small \textcolor{orange}{$2.05$}}		  
		\psfrag{2.1}[Bl][Bl]{\small \textcolor{green}{$2.1$}}		  
		\psfrag{2.2}[Bl][Bl]{\small \textcolor{cyan}{$2.2$}}		  
		\psfrag{2.4}[Bl][Bl]{\small \textcolor{blue}{$2.4$}}		  
		\psfrag{H2O}[cc][Bc]{\colorbox{white}{$\textrm{H}_2\textrm{0}$}}
		\includematlabfig{7cm}{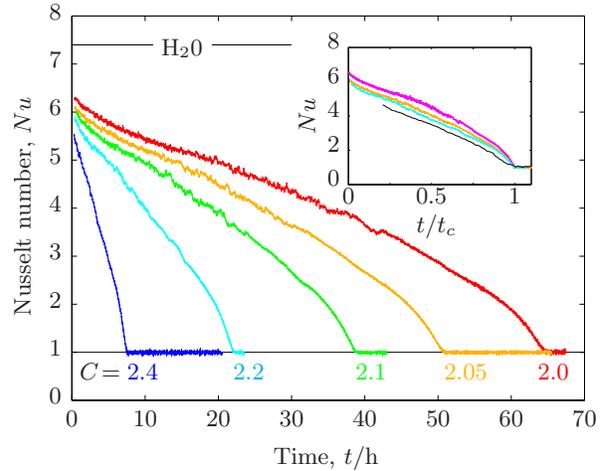}
	\end{minipage}	}
	\put(-11,45){
	\begin{minipage}{5cm}
		\footnotesize
		\xbottomlabel[2mm]{$t/t_c$}
		\yleftlabel[2mm]{$Nu$} 
		\includematlabfig{2.5cm}{Fig-NuC-inset.eps}
	\end{minipage}	}
\end{picture}
\end{center}
\caption{Aging of the Nusselt number: samples with different concentration $C$ (in $\SI{}{wt\%}$) are submitted to a $\Delta T = \SI{8}{K}$ temperature difference in convective configuration, and we plot the time evolution of the Nusselt number. Initially the heat transfer is efficient, close to the one reached with water in similar conditions, but the increasing viscosity of Laponite slows down the convective motion to eventually stop the Rayleigh B\'enard instability for a given time $t_c(C)$. In the inset, time axis is rescaled by $t_c$ for each ploted concentration (from top to bottom: $C= \SI{(1.8}{}$, $\SI{2.05}{}$, $\SI{2.2}{}$ and$\SI{2.6)}{wt\%}$). Although the curves are very similar, we notice a slight decreasing concavity of $Nu(t/t_c)$ with $C$.} \label{Fig-NuC}
\end{figure}

To study the influence on convection of the aging of Laponite preparations, let us focus on the time evolution of the heat transfer in a convective configuration: we heat the cell from below and impose a constant temperature difference $\Delta T = \SI{8}{K}$, and record the injected power. A simple ratio with the equivalent quantity in a conduction configuration (with necessary corrections to compensate for heat losses) define the Nusselt number $Nu$ of this sample. In figure \ref{Fig-NuC}, we present its time evolution for a few Laponite preparations with concentrations between $\SI{2.0}{wt\%}$ and $\SI{2.4}{wt\%}$. Initially, the solution has a very low viscosity (sligthly higher than water) and the heat transfer is efficient, leading to a Nusselt number close to the one reached with water in equivalent conditions: $Nu = \SI{7.4}{}$ (for $Ra=\SI{1.5e6}{}$). As the sample ages, its viscosity increases, reducing the Rayleigh number and thus the Nusselt number, up to the moment where convection disappear: the Rayleigh number is smaller than the critical one, and the heat transfer is purely conductive ($Nu=\SI{1}{}$). This instant define a characteristic time: the duration of convection for a specific sample will be denoted as $t_c$. Although the concave shapes of all the curves are quite similar, no simple rescaling was found in order to get a master curve: in the inset of figure \ref{Fig-NuC}, we present a simple renormalization of the time axis by $t_c$, evidencing the slight decreasing concavity of $Nu(t/t_c)$ with the concentration $C$.

\begin{figure}[!ht]  
\begin{center}
	\xbottomlabel{Concentration, $C/\SI{}{wt\%}$}
	\yleftlabel[5mm]{Convection duration, $t_c/\SI{}{h}$}
	\includematlabfig{7cm}{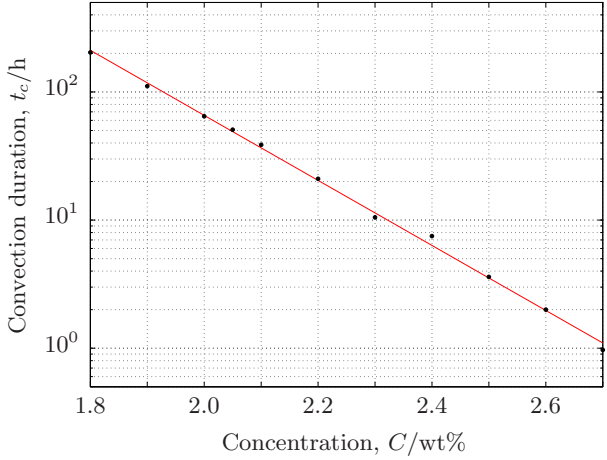}
\end{center}
\caption{Convection time $t_c$ as a function of sample concentration $C$. Measurements are all done with $\Delta T = \SI{8}{K}$.The vertical logarithmic scale underline that $t_c$ is a simple exponential function of $C$: $t_c=t_0 e^{-C/C_0}$ with $t_0=\SI{7.8e6}{h}$ and $C_0=\SI{0.17}{wt\%}$.} \label{Fig-t_c/C}
\end{figure}

In figure \ref{Fig-t_c/C} we plot the convection time $t_c$ as a function of sample concentration $C$, for $\Delta T = \SI{8}{K}$. $t_c$ turns out to be a simple exponential function of $C$: a best fitting procedure leads to
\begin{equation} \label{Eq-t_c(C)}
t_c=t_0 e^{-C/C_0}
\end{equation}
with $t_0=\SI{7.8e6}{h}$ and $C_0=\SI{0.17}{wt\%}$. This behavior of typical aging time versus concentration is consistent with previous observations on Laponite aging \cite{Kroon}, although coefficients $t_0$ and $C_0$ are quite different since the physical properties studied are very different. In order to perform several experiments in a reasonable time, but with a typical aging time much larger than the main thermal diffusion scale $\tau = h^2/\kappa$, we choose to work with $\SI{2.2}{wt\%}$ samples, for which $t_c \approx\SI{20}{h}$. This way, we can make one experiment per day and still have more than one order of magnitude between $t_c$ and $\tau =\SI{1.1}{h}$.

\begin{figure}[!ht] 
\unitlength=1mm
\begin{center}
\begin{picture}(0,73)(2.6,3)
	\put(-42.4,43){
	\begin{minipage}{\columnwidth}
		\xtoplabel{Time, $t/\SI{}{h}$}	
		\yleftlabel[2mm]{Nusselt number, $Nu$}
		\includematlabfig{7cm}{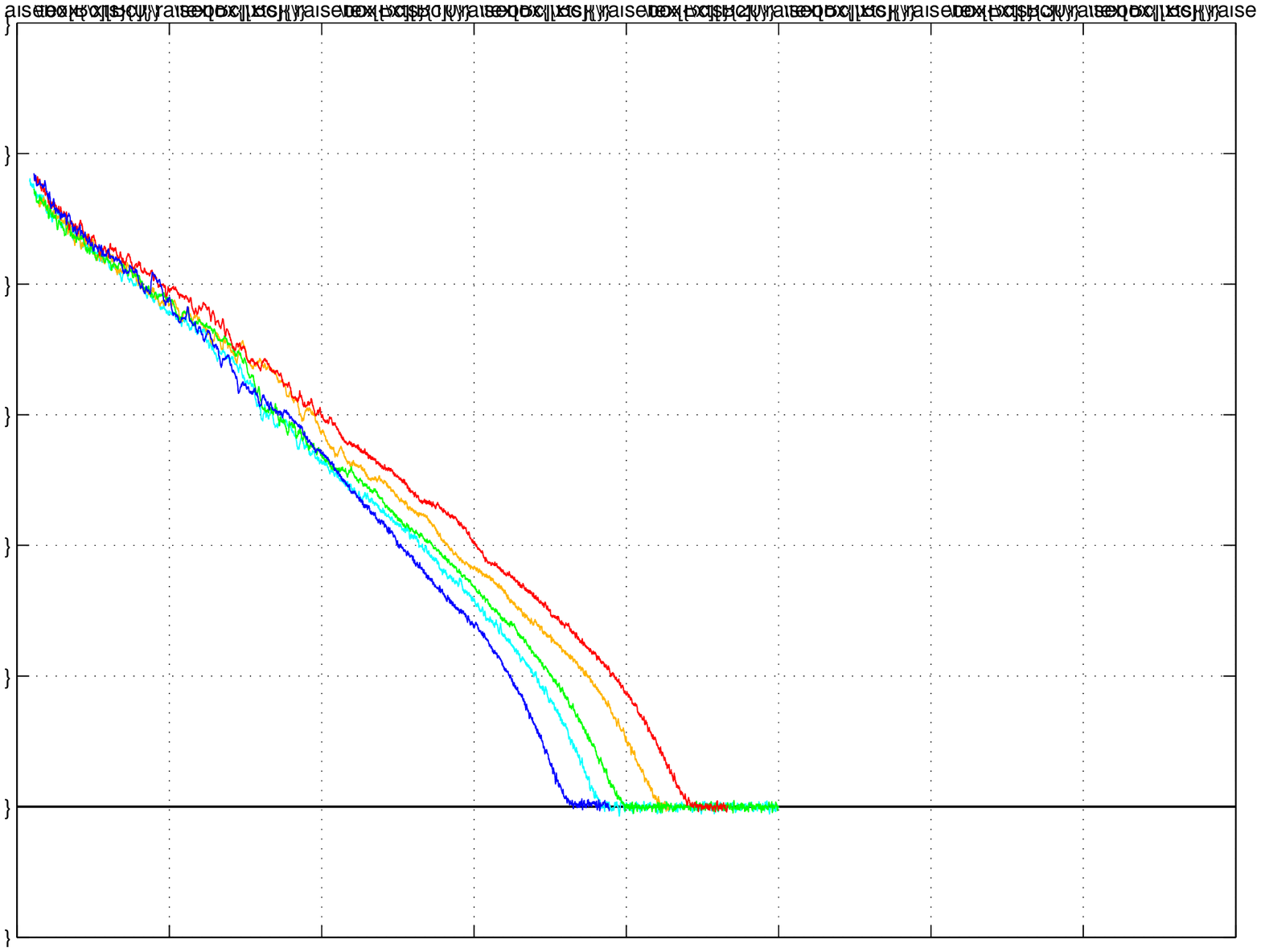}
	\end{minipage}	}
	\put(-11,27){\vector(1,0){10}}
	\put(2,17){\vector(0,1){10}}
	\put(13,27){\vector(-1,0){5}}
	\put(-10,54){\vector(-1,0){13}}
	\put(-28,20){\includegraphics[width=2.1cm]{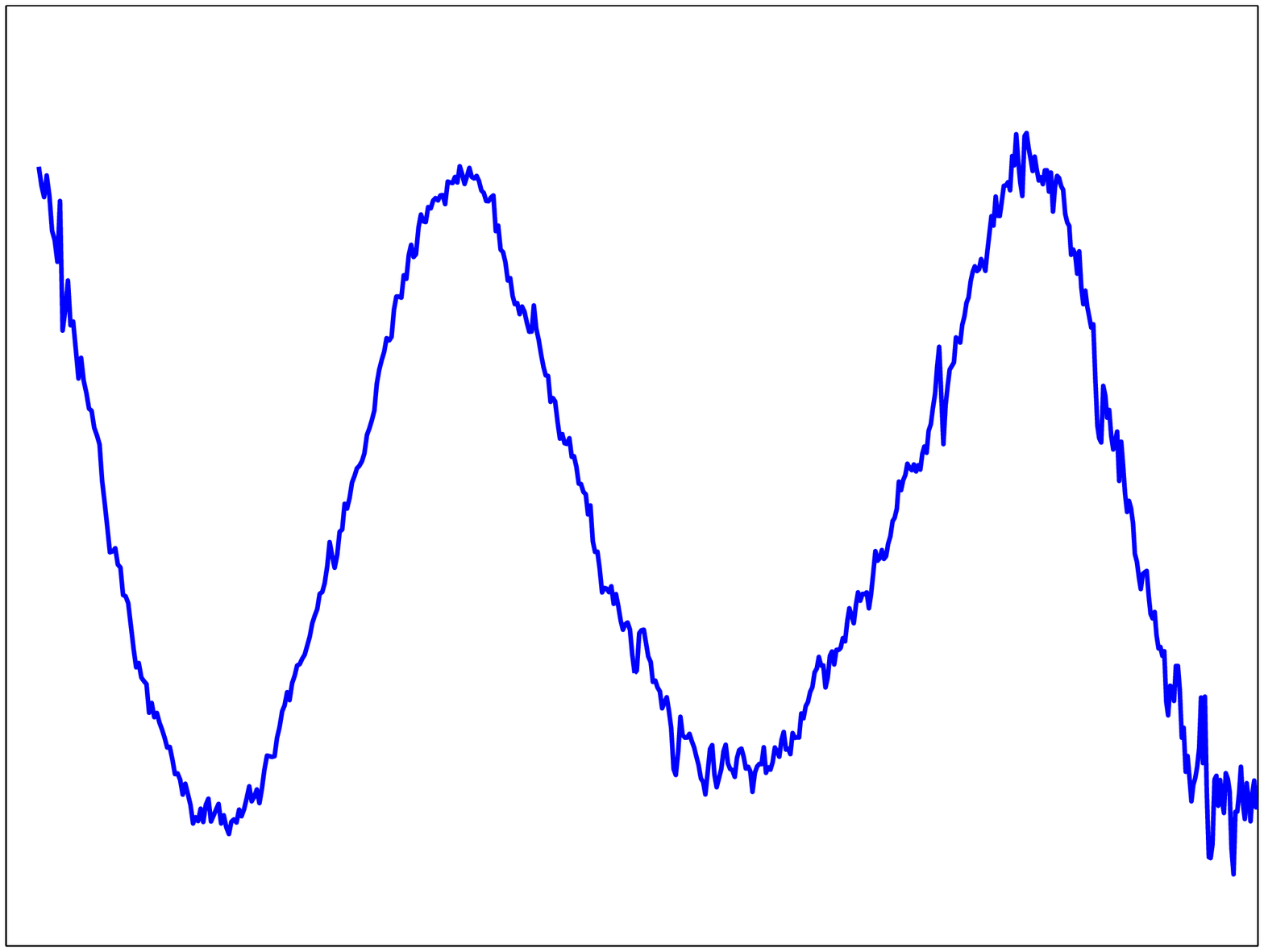}}
	\put(-27.8,20.5){\tiny \textcolor{blue}{5 rolls}}
	\put(-8,3){\includegraphics[width=2.1cm]{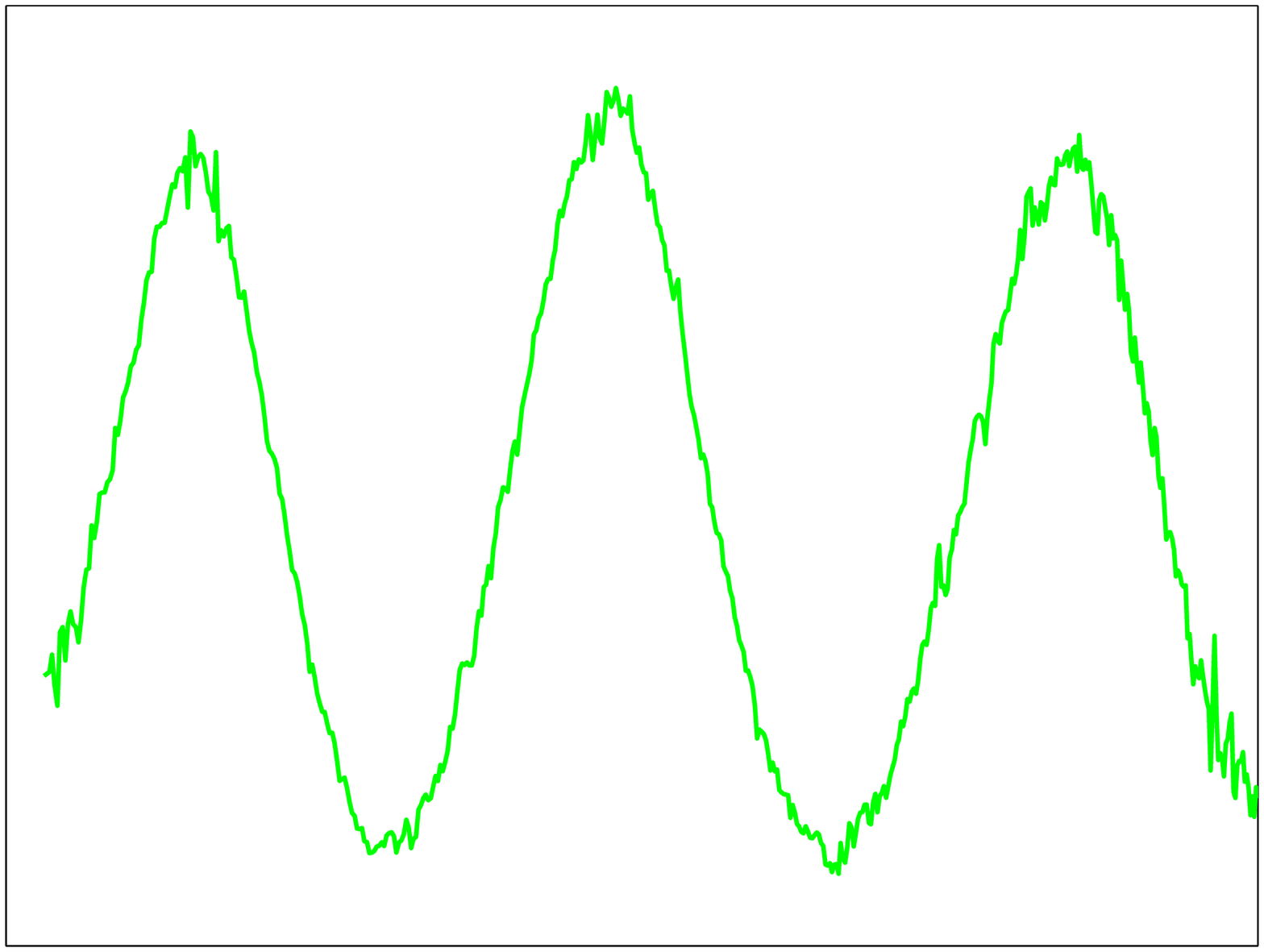}}
	\put(-7.7,3.5){\tiny \textcolor{green}{6 rolls}}
	\put(13,25){\includegraphics[width=2.1cm]{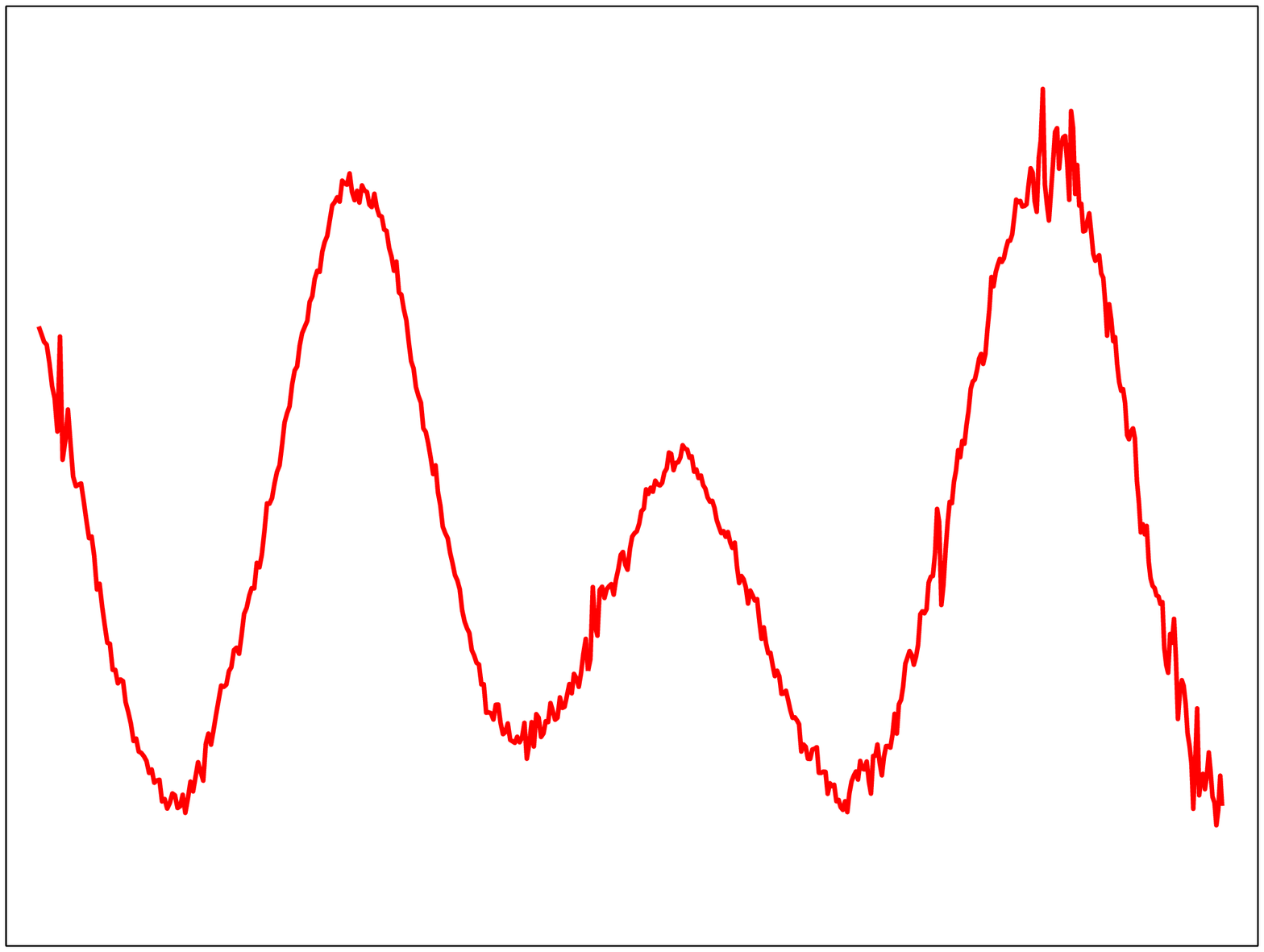}}	
	\put(27.2,25.4){\tiny \textcolor{red}{7 rolls}}
	\put(-13,47){\includegraphics[width=2.1cm]{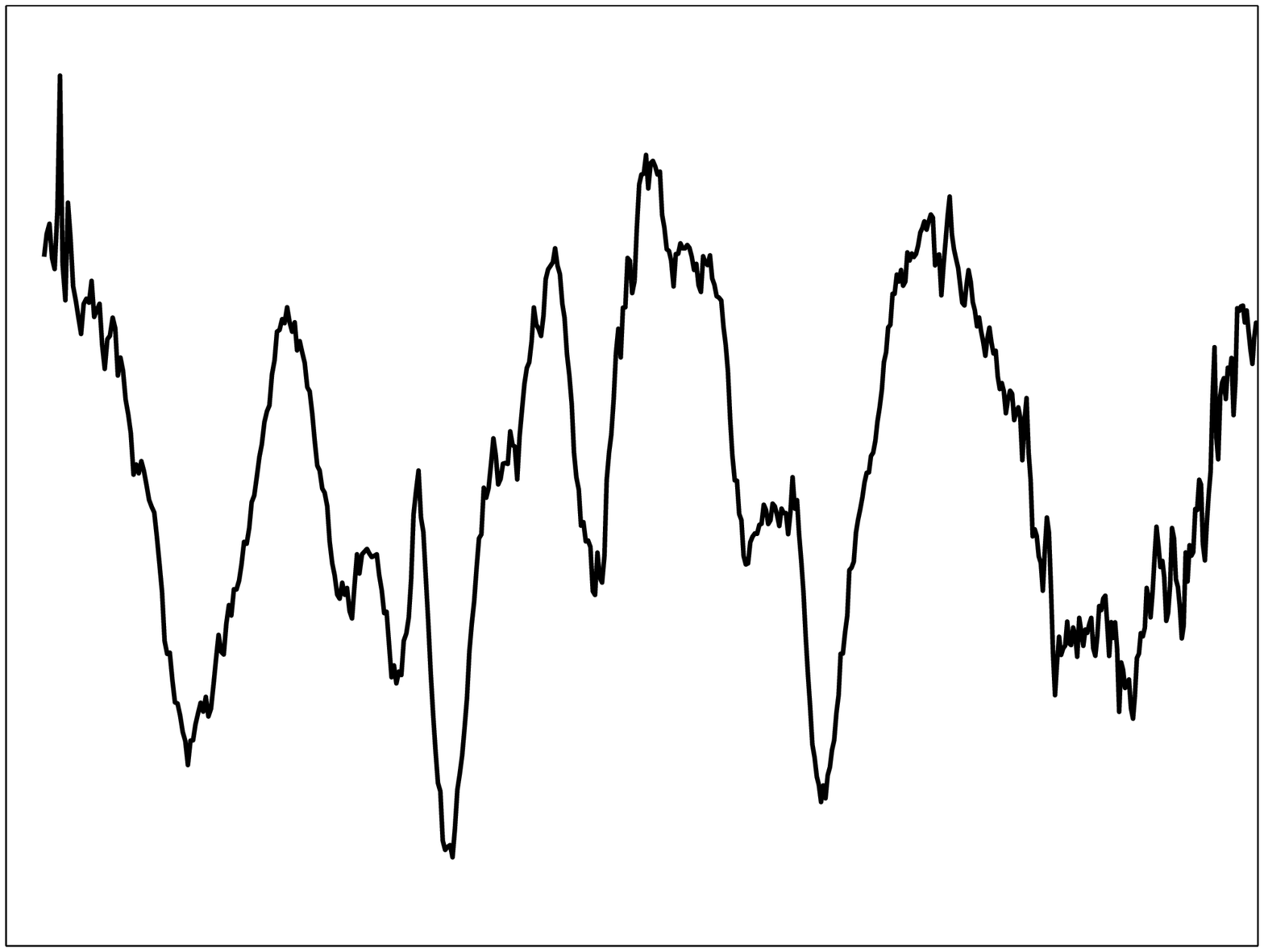}}
	\put(-10.7,61.2){\tiny turbulent pattern}
\end{picture}
\end{center}
\caption{Reproducibility of experiments: time evolution of Nusselt number, for $\Delta T= \SI{8}{K}$ and $C=\SI{2.2}{wt\%}$. As shown by the visualization technique, the convective flow is turbulent when all curves initially superpose. The final decaying of $Nu$ to $1$ is pattern dependent: $t_c$ is smaller for fewer rolls.} \label{Fig-NuPattern}
\end{figure}

If we repeat this same simple convection experiment with several samples of concentration $C=\SI{2.2}{wt\%}$, we see an important dispersion in $t_c$ : as shown in figure \ref{Fig-NuPattern}, $t_c$ varies between $\SI{18}{h}$ and $\SI{22}{h}$. As emphasized in the paragraph on Laponite preparation, we take great care to make every preparation in the same way, and the differences between these sample history are tiny. The precision on concentration for example, better than $\SI{0.002}{wt\%}$, exclude a interpretation of this dispersion in terms of sample composition : indeed, this imprecision leads to a $\SI{0.5}{h}$ incertitude according to eq.\ref{Eq-t_c(C)}. Moreover, we see that all the curves exactly superpose in a first regime, for the first $\SI{5}{h}$, and only then the behavior of each experiment changes to fall more or less quickly into the conductive state. The visualization of the pattern gives an interesting insight of the mechanism at work: the convective flow is turbulent initially, with small structures quickly evolving in time (the typical time of decorrelation of the pattern is $\SI{5}{s}$ after a $\SI{2}{h}$ aging). This state is reproducible among samples and leads to an efficient heat transfer. As the viscosity of the sample increases, the flow slowly evolves towards a more structured pattern, involving steady like convection rolls. The preferred pattern in a cell with $6:1$ aspect ratio should be of 6 steady convection rolls, but the lateral equilibration time of the cell is about $\SI{42}{h}$, longer than the existence time of the pattern! Therefore, other patterns may be initially selected by the turbulent flow, and can survive till the end of convection. This is what is observed in the last hours of convection of our samples: some patterns with 5,6 or 7 rolls have been observed, as well as other more complicated structures (generally unsteady - not shown). It turns out qualitatively that patterns with less rolls, fall quicker in a conductive state. The interpretation of the dispersion of the convection times $t_c$ would be the following: when smaller flow structures are implied, the average shear rate of the flow is higher, thus delaying the normal aging of the sample and maintaining convection for longer times.

\section{Discussions and conclusions} \label{Part-conclusion}

We present a study of low concentration Laponite preparations submitted to a destabilizing vertical temperature gradient. Except for rheological behavior, all the relevant properties of the samples are those of water, which in similar experimental conditions present turbulent convection ($Ra=\SI{1.5e6}{}$, $Nu=\SI{7.5}{}$). As measured with a standard rheometer, the complex viscosity of the samples evolves in time from fluid like to a solid like behavior. The onset of viscoelastic properties coincide with the increase of viscosity beyond the critical value for which convection disappear.

In the convection cell, this transition can be followed with the decaying of the Nusselt number from 6 to 1, as the heat transfer relaxes from a turbulent convective flow to a purely conductive state. The duration of the convective state $t_c$, depends strongly on the sample concentration : it decreases exponentially on more than 2 orders of magnitude when $C$ changes between $\SI{1.8}{wt\%}$ and $\SI{2.7}{wt\%}$. This sensitivity shows that convection is a very sensitive tool to investigate Laponite aging.

\begin{figure}[!ht]
	\xbottomlabel{Time, $t/\SI{}{h}$}
	\yleftlabel[3mm]{Mean shear rate, $\left<|\nabla v|^2\right>^{1/2}/\SI{}{s^{-1}}$} 
	\includematlabfig{7cm}{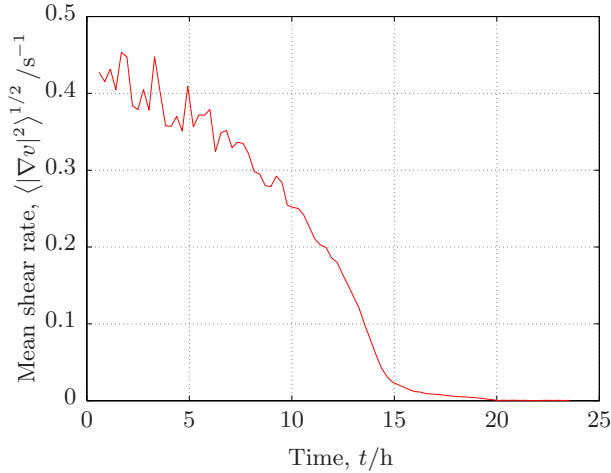}	
\caption{The mean shear rate to which the sample is submitted to is estimated from eq. \ref{Eq-shear} (see text for details). Although quite small, this small forcing associated to the thermally driven flow delays the aging of the sample, that is the formation of the colloidal glass.}  \label{Fig-shear_rate}
\end{figure}

For a given concentration, $t_{c}$ is longer than expected from rheological measurements : while $\nu^*$ crosses the critical viscosity $\nu_{c}$ around $\SI{15}{h}$, convection last about $\SI{20}{h}$. Moreover, $t_c$ presents an important dispersion, which is attributed to the Rayleigh B\'enard instability pattern: convection can be up to $\SI{20}{\%}$ longer when more rolls are implied. These effects can be interpreted considering the effect of the flow on aging : the small shear associated to the thermally driven fluid motion can delay aging, delaying the formation of the colloidal glass. Indeed, let us consider the global energy balance in the cell \cite{Koschmieder}:
\begin{equation} \label{Eq-shear}
\left<|\nabla v|^2\right> = \frac{\kappa^2}{h^4}Ra(Nu-1)
\end{equation}
The left term of this equation represent the space average of the velocity gradient, that is the mean shear rate to which the sample is submitted. Using the measurement of $\nu^*$ at $\SI{1}{Hz}$ (figure \ref{Fig-nu}) as an estimation for the kinematic viscosity, equation \ref{Eq-Ra} to compute the Rayleigh number, and the average behavior of the Nusselt number of figure \ref{Fig-NuPattern}, we can plot on figure \ref{Fig-shear_rate} a rough estimation of the time evolution of the shear rate to which the sample is submitted. The order of magnitude of this shear rate is $\SI{0.3}{s^{-1}}$. Although it is quite small, this forcing delays the aging of the preparation, and convection reveals this behavior by presenting a sensitivity to the flow pattern: more rolls (smaller spacial scale) present more shear for the sample, thus a longer convection state.

In other experiments performed on Laponite preparation to study their behavior under shear \cite{Coussot}, to get a stationary viscosity around $\SI{1}{Pa.s}$, the shear rate needed is about $\SI{25}{s^{-1}}$ for concentrations of $\SI{(3}{}$ to $\SI{3.5)}{wt\%}$. The concentrations we probe in our experiment are much \linebreak smaller, thus lowering the viscosity and probably the driving needed to get a stationary state. Anyway, our shear rate stand 2 orders of magnitude below this value, and we couldn't reach such a stationary state. Moreover, increasing the convective shear to larger values appears to be quite difficult: according to eq. \ref{Eq-shear}, it would mean increasing the Rayleigh and Nusselt numbers. $Nu$ is known to scale roughly as $Ra^{1/3} \propto h$ as long as $Ra>>Ra_{c}$, so that $\left<|\nabla v|^2\right>$ is independent on the value of $h$, and the only way to increase it is to raise $\Delta T$. We already use quite a high temperature difference ($\Delta T = \SI{8}{K}$), and cannot pretend to gain an order of magnitude on this parameter without loosing the Boussinesq approximation \cite{Koschmieder}. Thus the only chance to reach a stationary convective state would be to lower even more the concentration, which implies according to eq. \ref{Eq-t_c(C)} that we would have to deal with month (year \nolinebreak ?) like time scales...

As a conclusion, let us eventually try to answer the few questions we had in mind when beginning this study. First, thermal properties of low concentration Laponite preparation are not aging, they are in fact those of water within a few percent. Second, convective flow can indeed modify the aging behavior, delaying the formation of the colloidal glass when more convection rolls are implied. But anyway, it probably won't make a suitable driving method to study the fluctuation-dissipation ratio of a forced glassy system: we are unable to reach a stationary state which would ease the measurement of fluctuations.

However, this study gives a nice illustration of the coupling between flow and aging. It also sheds light on an interesting point: when forcing a soft glassy material, flow instabilities can play a significant role in the overall glassy behavior. In a wider field of view, this conclusion can be extended to many systems with inner processes (phase transitions, chemical reactions, etc.) occurring on time scales comparable to the flow ones.

{\noindent \bf Acknowledgments }

We acknowledge useful discussion with L. Berthier, A. Petrosyan, E. Freyssingeas, S. Ciliberto and B. Abou. We thank J.-F. Palierne for the rheological measurement facilities at the Laboratoire de Physique, ENS Lyon, France. Laponite RD \cite{Laponite} was gracefully provided by Laporte Absorbent LDTA. This work has been fully supported by FONDECYT Grant 3010067.

\end{document}

%% file: matlabfig.tex
\newlength{\xbs}
\newlength{\xts}
\newlength{\yls}
\newlength{\yrs}

\newlength{\figleftspace}
\newlength{\figrightspace}
\newlength{\figbottomspace}
\newlength{\figtopspace}

\newlength{\leftlabelwidth}
\newlength{\rightlabelwidth}

\newcommand{\toplabel}{}

\newcommand{\xtoplabel}[2][\baselineskip]{
	\renewcommand{\toplabel}{#2}
	\setlength{\figtopspace}{#1}
	}
	
\newsavebox{\bottomlabel}

\newcommand{\xbottomlabel}[2][\baselineskip]{
	\sbox{\bottomlabel}{#2}
	\setlength{\figbottomspace}{#1}
	}
	
\newsavebox{\leftlabel}

\newcommand{\yleftlabel}[2][5mm]{
	\sbox{\leftlabel}{\rotatebox{90}{#2}}
	\setlength{\figleftspace}{#1}
	\settowidth{\leftlabelwidth}{\usebox{\leftlabel}}
	}
	
\newsavebox{\rightlabel}

\newcommand{\yrightlabel}[2][5mm]{
	\sbox{\rightlabel}{\rotatebox{90}{#2}}
	\setlength{\figrightspace}{#1}
	\settowidth{\rightlabelwidth}{\usebox{\rightlabel}}
	}
	
\newcommand{\includematlabfig}[2]{
	\psfragscanon
	\begin{center}
	\null \hspace*{\figleftspace} \hspace*{\leftlabelwidth} \toplabel
	\hspace*{\rightlabelwidth} \hspace*{\figrightspace} \null
	\vspace{\figtopspace}
	
	\begin{minipage}{\leftlabelwidth}
		\usebox{\leftlabel} 
	\end{minipage}
	\hspace{\figleftspace}
	\begin{minipage}{#1}
		\includegraphics[width=#1]{#2}
	\end{minipage}
	\hspace{\figrightspace}
	\begin{minipage}{\rightlabelwidth}
		\usebox{\rightlabel}
	\end{minipage}
	
	\vspace{\figbottomspace}
	\null \hspace*{\figleftspace} \hspace*{\leftlabelwidth} \usebox{\bottomlabel}  	\hspace*{\rightlabelwidth} \hspace*{\figrightspace} \null
	\end{center}
	
	\yleftlabel{0mm}{}
	\yrightlabel{0mm}{}
	\xtoplabel{0mm}{}
	\xbottomlabel{0mm}{}
}

%% file: LaponiteConvection.bbl
\begin{thebibliography}{}
\bibitem{Laponite} Laponite RD is a registered trademark of Laporte
Absorbents, P.O Box 2, Cheshire, UK, which gracefully provided the
material needed for this study.
\end{thebibliography}
